%% file: arxivISIT26.tex
\documentclass[conference]{IEEEtran}
\IEEEoverridecommandlockouts
\usepackage{algorithm,algorithmic,amsmath,amssymb,amsthm,bbm,caption,cite,color,comment,graphicx,microtype,setspace,subcaption,tabularx,url}
\usepackage[USenglish]{babel}
\usepackage[T1]{fontenc}
\usepackage[utf8]{inputenc}
\usepackage{hyperref}
\usepackage{mathtools}
\usepackage{graphicx}
\usepackage{subcaption}
\usepackage{pgfplots}
\usepackage{algorithm,algorithmic,amsmath,amssymb,amsthm,bbm,cite,color,graphicx,microtype,url}
\usepackage[USenglish]{babel}
\usepackage[utf8]{inputenc}
\usepackage[T1]{fontenc}
\usepackage{hyperref}
\usepackage{subcaption}
\usepackage{tikz}
\usepackage{tikzscale}
\usepackage{babel}
\usepackage{anyfontsize}
\usepackage{balance}

\usepackage{tikz,pgfplots}
\usetikzlibrary{arrows, arrows.meta, backgrounds, calc, intersections, decorations.markings, decorations.pathreplacing, positioning, shapes}
\pgfplotsset{compat=1.17}

\usepackage{caption}
\usepgfplotslibrary{fillbetween}
\usepgfplotslibrary{colormaps}
\usetikzlibrary{positioning,3d}
\pgfplotsset{compat=newest}

\usepackage{enumitem}

\usepackage{pgfplots}
\pgfplotsset{compat=newest}

\usepackage{tikz}
\usetikzlibrary{arrows,backgrounds,calc,intersections,decorations.pathreplacing,positioning,shapes}
 \usepackage{titlesec}
 \let\subparagraph\relax
 \titlespacing{\section}{0pt}{5pt plus 2pt minus 1pt}{3pt plus 1pt minus 0pt}
 \titlespacing{\subsection}{0pt}{4pt plus 2pt minus 1pt}{2pt plus 1pt minus 0pt}
\usepackage[figurename=Fig.,font=small]{caption}

{\newtheorem{theorem}{Theorem}}
{}
{}
{\newtheorem{proposition}{Proposition}}
{}
{}
{\newtheorem{corollary}{Corollary}}

\input{./symbols.tex}

\title{The Effect of Noise Correlation on MMSE Channel Estimation in One-Bit Quantized Systems}
\author{
\IEEEauthorblockN{Minhua~Ding, Prathapasinghe~Dharmawansa, Italo~Atzeni, and  Antti~Tölli \\ Centre for Wireless Communications, University of Oulu, Finland\\
E-mail: \{minhua.ding, prathapasinghe.kaluwadevage, italo.atzeni, antti.tolli\}@oulu.fi}
\thanks{This work was supported by the Research Council of Finland (336449 Profi6, 348396 HIGH-6G, 357504 EETCAMD, and 369116 6G~Flagship).} \vspace{-1mm}}

\begin{document}

\maketitle
\begin{abstract}
This paper analyzes the impact of spatially correlated additive noise on the minimum mean-square error (MMSE) estimation of multiple-input multiple-output (MIMO) channels from one-bit quantized observations. Although additive noise can be correlated in practical scenarios, e.g., due to jamming, clutter, or other external disturbances, the effect of such correlation on the MMSE channel estimator in this setting remains unexplored in prior work. Against this backdrop, we derive a novel analytical expression for the general MIMO MMSE  channel estimator, which is inherently nonlinear in one-bit observations, and accommodates arbitrary channel and noise correlation structures. To further characterize the impact of noise correlation, we subsequently specialize the general MMSE expression to certain tractable multi-antenna configurations in which both the channel and the noise assume single-parameter constant correlation structures. Our analyses reveal nontrivial, noise‑correlation‑induced scenarios in which the estimator remains linear despite non‑zero channel and noise correlation parameters.
Moreover, the results indicate that, at low-to-medium signal-to-noise ratio, noise correlation improves the MMSE performance when channels are uncorrelated, but degrades performance when channels are strongly correlated.
\end{abstract}

\section{Introduction}
\label{sec:intro}
Massive multiple-input multiple-output (MIMO) systems are widely regarded as a cornerstone technology for enabling next-generation high-data-rate and ultra-reliable wireless communications \cite{Emil-etal-2017, Wu_YF_Liu2024,Bernardo_IT2022,Yang_Combes_IT25,Y_Li_et_al_BLMMSE,  J_Mo_R_Heath}. However, the significant power consumption associated with fully digital architectures employing high-resolution analog-to-digital converters (ADCs)
has motivated extensive research into low-resolution alternatives. Among these, systems with one-bit ADCs have received particular attention due to their potential for drastically reduced power consumption and hardware complexity \cite{Yang_Combes_IT25, Dabeer_Masry_IT08,Y_Li_et_al_BLMMSE, Choi_ML,Ding_et_al_2025,J_Mo_R_Heath}. 

Several key aspects of one-bit quantized multi-antenna systems have been studied in the literature, including capacity and its bounds \cite{J_Mo_R_Heath, Yang_Combes_IT25}, receiver design and data detection \cite{Wen_et_al, Choi_ML}, and channel estimation \cite{Y_Li_et_al_BLMMSE, Fesl_etal_TSP2023,Nguyen_DL2023, Ding_et_al_2025, Choi_ML,Studer2016, Wen_et_al}. 
 Accurate channel estimation is particularly crucial for enabling coherent communication. 
Various estimators have been examined for this setting, such as the maximum likelihood (ML), maximum a posteriori (MAP),  and minimum mean-square error (MMSE) estimators \cite{Wen_et_al, Fesl_etal_TSP2023, Ding_et_al_2025}. A notable limitation of prior work is the assumption of additive white Gaussian noise (AWGN), and thus the impact of correlated noise on these estimators remains largely unaddressed.

In multi-antenna systems, noise correlation across antenna elements arises due to mutual coupling, external interference, or other disturbances \cite{Charmain_Prathapa_TIT, Neh03, Stoica_correlated_noise_field, Besson_2023, Craeye_2005, Sharma_2013_1, Sharma_TWC2013}. Moreover, in interference-limited cellular environments, aggregate co-channel interference often dominates thermal noise, resulting in a non-white disturbance that defines the effective signal-to-interference-plus-noise ratio (SINR) \cite{Shah-Haimovich98}. While the effect of noise correlation has been partially examined in the context of capacity bounds for quantized MIMO channels \cite{Mezghani_Nossek}, its specific impact on channel estimation performance remains unexplored.

Motivated by this research gap, we analyze the effect of noise correlation on the MMSE channel estimator in multi-antenna systems with one-bit quantization at the receiver. The main contributions of this work are summarized as follows.

\begin{enumerate}
    \item We derive an analytical expression for the {\it generally nonlinear} MIMO MMSE channel estimator with one-bit observations that explicitly incorporates noise correlation. This analytical expression inherently involves high-dimensional orthant probabilities. Our result holds for general channel and noise correlation models beyond the classical Kronecker structure~\cite{Hanlen_Grant_TIT12,Jafar_Goldsmith_ISIT03, Wagner_IT08} and subsumes the AWGN case previously studied in the literature (see, e.g.,  \cite{Fesl_etal_TSP2023,Ding_et_al_2025} and references therein). 

    \item To obtain tractable insights, the general result is specialized to single‑parameter constant correlation models (both real and complex) for both the channel and noise~\cite{V_Aalo, Hyundong_Shin_Lee_IT2003, Chen_Chintha1_equal_corr_model2, Chen_Chintha2_equal_corr_model3, Beaulieu_generalized_equal_corr_IT}, yielding compact closed-form expressions for certain single-input multiple-output (SIMO) configurations~\cite{Hyundong_Shin_Lee_IT2003}. 
    Specifically, we consider SIMO systems with two receive antennas under the complex correlation model, and systems with an arbitrary number of receive antennas under the real correlation model. 

    \item 
    The analyses reveal  nontrivial relationships among the channel correlation, noise correlation, and signal-to-noise ratio (SNR) under which the MMSE estimator remains linear even when both channel and noise correlation parameters are non‑zero. 
    To the best of our knowledge, this linearity phenomenon under correlated conditions has not been reported previously. Our results further indicate that noise correlation enhances low-to-medium SNR estimation performance for uncorrelated channels but degrades 
    it when channels are highly correlated.
\end{enumerate}
\emph{Notation}. The diagonal matrix with  diagonal entries given by the vector $\mathbf{a}$ is denoted by $\diag{\mathbf{a}}$, whereas $\diag{\mathbf{C}}$ is diagonal and shares  the same diagonal entries as the matrix $\mathbf{C}$.
$\sgn(\cdot)$ denotes the signum function. $\real{\cdot}$ and $\imag{\cdot}$ denote the real and imaginary parts, respectively. The imaginary unit is $j=\sqrt{-1}$.  $\mathcal{CN}_p(\a, \mathbf{D})$  (resp. $\mathcal{N}_p(\a, \mathbf{D})$) denotes a complex-valued (resp. real-valued) $p$-dimensional multivariate  Gaussian distribution with mean vector $\a$  and covariance matrix $\mathbf{D}$. $(\cdot)\Trans$, $(\cdot)\Herm$ and  $(\cdot)^*$ denote transpose, Hermitian transpose and complex conjugate, respectively.  $\I_p$ is the $p\times p$ identity matrix and $\mathbf{1}_n= (1\;\ldots\;1)^T\in\mathbb{R}^n$. $\Prob{\cdot}$ denotes  probability and $\myexp{\cdot}$  mathematical expectation. $\mathbb{E}_X\{\cdot\}$ denotes expectation with respect to $X$.    
$\vec(\cdot)$ denotes vectorization and $\|\cdot\|$ the Euclidean vector norm. $|\cdot|$ denotes the modulus of a complex scalar or the determinant of a square matrix. 
$Q(x)=\frac{1}{\sqrt{2\pi}}\int_x^\infty e^{-t^2/2} {\rm d}t$  is the Gaussian-Q function. 
\section{System Model and Estimation Problem}
\subsection{System Model}\label{sec: sys_model_model}
Consider a  Rayleigh block-fading~\cite{Hassibi_Hochwald_IT03}  MIMO channel $\mathbf{H}\in\mathbb{C}^{n\times m}$ with $m$ transmit antennas and $n$ receive antennas. For channel estimation, a pilot matrix $\mathbf{S}\in\mathbb{C}^{m\times\tau}$  is sent column-wise over $\tau$ time slots by the transmitter. At the receiver side,  over $\tau$ time slots,  the received signal vectors form a matrix $\mathbf{Y}\in\mathbb{C}^{n\times\tau}$, which is given by 
\begin{align}
    \mathbf{Y} =\mathbf{H}\mathbf{S} +\mathbf{N},
     \label{eqn: signal_model_matrix_form} 
\end{align}
where the noise matrix $\mathbf{N}\in\mathbb{C}^{n\times\tau}$  collects the $n$-dimensional Gaussian noise vectors over $\tau$ time slots. The observable signal matrix $\mathbf{R}$ is the one-bit quantized $\Y$, i.e., 
\begin{align}
    \mathbf{R} = \quantoneb{\mathbf{Y}}, \label{eqn: observable_R_after_quant}
\end{align}
where \(\quantoneb{\cdot} =\sgn(\real{\cdot})+j\sgn(\imag{\cdot})\) denotes the element-wise memory-less one-bit quantization of the complex-valued argument.
Let $\y=\vec(\mathbf{Y})\in\mathbb{C}^{\tau n}$, $\h=\vec(\mathbf{H})\in\mathbb{C}^{mn}$, $\n=\vec(\mathbf{N})\in\mathbb{C}^{\tau n}$, and $\r =\vec(\mathbf{R})\in\mathbb{C}^{\tau n}$. 
Then, \eqref{eqn: signal_model_matrix_form}  can be equivalently expressed as 
\begin{align}
    \y = (\mathbf{S}\Trans\otimes\mathbf{I}_{n})\h + \mathbf{n}, 
\label{eqn: system_model_UQ_vec}
\end{align}
where $\otimes$ denotes the Kronecker product. From   
\eqref{eqn: observable_R_after_quant}, we have
\begin{align}
    \r =\quantoneb{\y} \in\{\pm1\pm j\}^{\tau n}.
\end{align}
We assume that $\mathbf{n}\sim\mathcal{CN}_{\tau n}(\0, \boldsymbol{\Xi})$ and  
$\mathbf{h}\sim \mathcal{CN}_{mn}(\0, \boldsymbol{\Phi})$ are independent. The corresponding covariance  matrices $\boldsymbol{\Xi}\in \mathbb{C}^{\tau n \times \tau n}$ and $\boldsymbol{\Phi}\in\mathbb{C}^{mn\times mn}$ are  assumed to be known and positive definite. Therefore, we have $\y\sim \mathcal{CN}_{\tau n}(\0, \bmat)$ with 
\begin{align}
\bmat& 
=(\mathbf{S}\Trans\otimes \I_{n})\boldsymbol{\Phi}(\mathbf{S}^*\otimes\I_{n}) +\boldsymbol{\Xi}.  
\label{eq: B_mat}
\end{align}
    Note  that the general covariance models here  subsume  the widely used Kronecker models~\cite{Hanlen_Grant_TIT12,Jafar_Goldsmith_ISIT03, Wagner_IT08} as special cases. 

\subsection{Problem Statement}
Given the one-bit quantized observation vector $\r$ as well as the system parameters $\mathbf{S}$, $\boldsymbol{\Phi}$, and $\boldsymbol{\Xi}$,  we aim at obtaining an explicit analytical  expression for the  MMSE channel estimator in the above setting. To this end, following~\cite{S-M-Kay_estimation, Curry_R_Est_Control_quant}, the MMSE estimator can be written as
\begin{align}
    \esthmmse(\r) = \myexp{\h|\r}=\displaystyle\int_{\mathbb{C}^{m n}}\h f(\h|\r)\intd\h, \label{eqn: h_MMSE_definition}
\end{align}
 with the corresponding MMSE {\it per dimension} given by
\begin{align}
\mathrm{MMSE}&=\frac{1}{mn}\mathbb{E}\left\{\|\h-\esthmmse(\r)\|^2\right\}. 
 \label{eq: MMSE_general_simplified}
 \end{align}
 Here, $\myexp{\h|\r}$ denotes the conditional mean  and $f(\h|\r)$  is the conditional probability density function.

Although the estimator in \eqref{eqn: h_MMSE_definition} was analyzed in~\cite{Ding_et_al_2025} under the assumption of white noise (i.e., with $\boldsymbol{\Xi}=\I_{\tau n}$), the techniques employed therein do not extend to scenarios with correlated noise (c.f.~\cite[App. B]{Ding_et_al_2025}). Motivated by this, here we employ a new technique to derive a more general analytical expression for the MMSE channel estimator that incorporates noise correlation. Subsequently, we apply these general results to SIMO systems with specific channel and noise correlation structures, and quantify the impact of noise correlation on the corresponding MMSE estimators.
\section{The General MIMO MMSE Channel Estimator with Noise Correlation}
\label{sec:general results}
In this section, we present an expression for the general MIMO MMSE channel estimator with noise correlation. It turns out that this general MMSE expression is closely related to the  orthant probability of multivariate normal distributions~\cite{Childs-1967, Abrahamsom_64, Sinn_Keller_cov_of_zero_crossing, Tong_MVN_book}. Therefore, following \cite{Childs-1967, Abrahamsom_64, Sinn_Keller_cov_of_zero_crossing}, we define the orthant probability associated with $\boldsymbol{w}\sim \mathcal{N}_p(\0, \mathbf{\boldsymbol{\Omega}})$ as \[\OrthP{\boldsymbol{\Omega}^{-1}}=\frac{1}{(2\pi)^{\frac{p}{2}}\!\left|\boldsymbol{\Omega}\right|^{\frac{1}{2}}}
    \int_{\mathbb{R}_+^p} e^{-\frac{1}{2}\boldsymbol{w}\Trans\boldsymbol{\Omega}^{-1}\boldsymbol{w}} \intd\boldsymbol{w},\]
where $\boldsymbol{\Omega}\in\mathbb{R}^{p\times p}$ is positive definite and $\mathbb{R}_+^p$ denotes the non-negative orthant. The following theorem gives the exact general  MMSE channel estimator. The proof is omitted due to space limits.
\begin{theorem}\label{theo:general}
The MIMO MMSE channel estimator in the presence of noise correlation is given by
\begin{align*}
    \esthmmse(\r)   &=\frac{1}{2\sqrt{\pi}}\boldsymbol{\Phi}(\mathbf{S}^*\otimes\mathbf{I}_{n})(\diag{\mathbf{B}})^{-\frac{1}{2}}\frac{\mathbf{g}(\r, \bmatext)}{\OrthP{\boldsymbol{\Lambda}_{\r}
\bmatext^{-1}\boldsymbol{\Lambda}_{\r}}}, 
 \end{align*}
 where, for $k=1, \ldots, \tau n$,  the $k$-th element of the vector $\mathbf{g}(\r, \bmatext)\in\mathbb{C}^{\tau n}$ is given by
\begin{align}
 g_k(\r, \bmatext)&=\real{r_k}\OrthP{\mathbf{E}_k\boldsymbol{\Lambda}_{\r}
\bmatext^{-1}\boldsymbol{\Lambda}_{\r}\mathbf{E}_k\Trans}\nonumber
\\&\quad +j\imag{r_k}\OrthP{\mathbf{E}_{k+\tau n}\boldsymbol{\Lambda}_{\r}
\bmatext^{-1}\boldsymbol{\Lambda}_{\r}\mathbf{E}_{k+\tau n}\Trans}.
\label{eq: h_mmse_g_rB}
\end{align}
Furthermore, for $i=1, \ldots, 2\tau n$, $\mathbf{E}_i\in\mathbb{R}^{(2\tau n-1)\times (2\tau n)}$ denotes the matrix obtained by removing the $i$-th row of $\mathbf{I}_{2\tau n}$, $\boldsymbol{\Lambda}_{\r}=\diag{\begin{bmatrix}
 \real{\r}\\\imag{\r}   
\end{bmatrix}}$
is a diagonal matrix containing   all the observations (i.e., $\real{\r}, \imag{\r}\in\{\pm 1\}^{\tau n}$), and $\bmatext$ is given by \begin{align}
    \bmatext = \frac{1}{2}\begin{bmatrix}
        \real{\mathbf{B}}   & -\imag{\mathbf{B}}
        \\\imag{\mathbf{B}}  &\phantom{+}\real{\mathbf{B}}
    \end{bmatrix}.\label{eq: B_extended_definition}
\end{align}
\end{theorem}
The general MMSE estimator in Theorem~\ref{theo:general} accommodates arbitrary correlation structures beyond the commonly used Kronecker model~\cite{Hanlen_Grant_TIT12,Jafar_Goldsmith_ISIT03, Wagner_IT08}. This further highlights the utility of our expression in various applications involving non-trivial correlation structures. Moreover, Theorem~\ref{theo:general} shows that, in general, the MMSE estimator is a nonlinear function of the one-bit quantized observation vector $\mathbf{r}$.
 This nonlinearity persists except in special cases, as detailed in Corollary~\ref{coro: linearity} below.
It is also worth noting that, for $\boldsymbol{\Xi}=\mathbf{I}_{\tau n}$ (i.e., white noise), the above MMSE expression degenerates into \cite[(25)--(26)]{Ding_et_al_2025}.
\begin{corollary}\label{coro: linearity}
    Under the necessary and sufficient condition that each row (resp. each column) of $\bmatext$ in \eqref{eq: B_extended_definition} contains at most two nonzero elements, the MMSE estimator in Theorem \ref{theo:general} admits the linear form
    \begin{align}
    \esthmmse(\r)   &\!=\!\sqrt{\frac{\pi}{4}}\boldsymbol{\Phi}(\mathbf{S}^*\otimes\mathbf{I}_{n})(\diag{\mathbf{B}})^{-\frac{1}{2}}\mathbf{B}_{\rm sa}^{-1} \r, 
    \label{eq: corollary_linear_h_MMSE}
 \end{align}
    where \[\mathbf{B}_{\rm sa}=\arcsin\big((\diag{\mathbf{B}})^{-\frac{1}{2}}\mathbf{B}(\diag{\mathbf{B}})^{-\frac{1}{2}}\big)\]
    and  the $\arcsin$ function is applied element-wise to the real and imaginary parts separately. 
\end{corollary}
\begin{IEEEproof}
    The proof follows  similar arguments as for  \cite[Theorem 1]{Ding_et_al_2025} and is thus omitted.
\end{IEEEproof}
 
 Note that, with noise correlation incorporated, the linear estimator in the form given by \eqref{eq: corollary_linear_h_MMSE} generalizes the well-known Bussgang linear MMSE (BLMMSE) channel estimator proposed  in~\cite{Y_Li_et_al_BLMMSE} assuming AWGN. Corollary~\ref{coro: linearity} specifies the precise condition under which  the generalized MMSE channel estimator with noise correlation reduces to the linear form. 

Despite the broad utility of the general MMSE estimator derived above, the exact evaluation is limited by the numerical complexity of computing the high-dimensional orthant probability for a generic covariance matrix\cite{Childs-1967, Abrahamsom_64, Tong_MVN_book}. This is further highlighted by the fact that closed-form orthant probabilities associated with $\boldsymbol{w}\sim \mathcal{N}_p(\0, \mathbf{\boldsymbol{\Omega}})$ were reported in the literature only for very special cases, e.g.,  for $p\leq 3$ with arbitrary positive-definite $\boldsymbol{\Omega}\in \mathbb{R}^{p\times p}$ or for any diagonal positive-definite $\boldsymbol{\Omega}\in\mathbb{R}^{p\times p}$~\cite{Childs-1967, Abrahamsom_64, Sinn_Keller_cov_of_zero_crossing}. Therefore, in the sequel, we focus on tractable SIMO configurations to gain insights into the impact of noise correlation. 
\section{Effect of Noise Correlation on SIMO Systems} 
In this section, we focus on analytical characterization of the effect of noise correlation on the MMSE channel estimation performance in SIMO systems with one scalar pilot symbol $\sqrt{\gamma}s\in\mathbb{C}$ ($\gamma>0$). Here, we assume that $|s|^2=1$, i.e., $s$ is drawn from a unit-energy constant-envelope modulation scheme such as $M$-ary phase-shift keying, with $\gamma$ denoting the transmit power. 
Under the above assumptions, \eqref{eqn: system_model_UQ_vec} specializes to
\begin{align}
    \y = \sqrt{\gamma}s\h +\n \in\mathbb{C}^n, \label{eq: SIMO_UM_system_model}
\end{align}
with the one-bit outputs given by $ \r=\quantoneb{\y}$ and the associated matrix $\mathbf{B}$ (c.f. \eqref{eq: B_mat}) taking the form
\begin{align}
\mathbf{B}=\gamma\boldsymbol{\Phi}+\boldsymbol{\Xi} \in\mathbb{C}^{n\times n}. 
\label{eq: B_mat_SIMO}
\end{align}
To facilitate further analyses, we restrict our attention to a certain class of correlation matrices in which each member can be parameterized by a single scalar parameter\cite{V_Aalo, Hyundong_Shin_Lee_IT2003,Chen_Chintha1_equal_corr_model2, Chen_Chintha2_equal_corr_model3, Beaulieu_generalized_equal_corr_IT}. More specifically,  we employ the constant {\it complex correlation} structure given by
\begin{align}
[\boldsymbol{\Phi}]_{k,\ell}\!=\!\left\{
   \begin{array}{cc}\!\!
        \phi^*&  \text{if $\ell>k$}\\\!\!
        1& \text{if $\ell=k$} \\\!\!
        \phi &\text{if $\ell<k$}
   \end{array},\right.\;\;
   [\boldsymbol{\Xi}]_{k,\ell}\!=\!\left\{
   \begin{array}{cc}
       \!\! \xi^*&  \text{if $\ell>k$}\\\!\!
        1& \text{if $\ell=k$} \\\!\!
        \xi &\text{if $\ell<k$}
   \end{array},\right. \label{eq: Phi_Xi_equal_corr}
\end{align}
with $|\phi|<1$, $|\xi|<1$ and the corresponding {\it real correlation} matrices of the form
\begin{align}
\boldsymbol{\Phi}&=(1-\phi)\I_{n} +\phi \1_{n} \1_{n}\Trans\;,\; 
        \boldsymbol{\Xi}=(1-\xi)\I_{n}  +\xi \1_{n} \1_{n}^T,  \label{eq: real_Phi_Xi_equal_corr}
\end{align}
 with $\phi,\xi\in(-1/(n-1),1)$.
Accordingly, with a slight abuse of notation, the MMSE {\it per dimension}  in \eqref{eq: MMSE_general_simplified} (hereafter simply referred to as the MMSE) can be expressed as 
\begin{align}
\mathrm{MMSE}_n(\gamma, \phi, \xi)
&=\left(\trace(\boldsymbol{\Phi})-\mathbb{E}\left\{\|\esthmmse(\r)\|_2^2\right\}\right)/n\nonumber
\\&=1-\frac{1}{n}\mathbb{E}\left\{\|\esthmmse(\r)\|_2^2\right\}. \label{eq: MMSE_SIMO_general}
\end{align}
Clearly, from the above, 
the parameter $\gamma$ can be interpreted as the effective transmit SNR of the system under consideration. 
Below we analytically characterize the influence of noise correlation on the structure and performance of the MMSE channel estimator.
\subsection{$n=2$  with Complex Correlation Structures}
\label{sec: 2x2_cov_complex}
Under the setting with $n=2$, following (\ref{eq: Phi_Xi_equal_corr}), we obtain $\boldsymbol{\Phi}\!=\!\!\begin{bmatrix}
        1 & \phi^*\\\phi & 1
    \end{bmatrix}$ ($|\phi|<1$),  $\boldsymbol{\Xi}\!=\!\! \begin{bmatrix}
        1 & \xi^*\\\xi & 1
    \end{bmatrix}$ ($|\xi|
    <1$), and \[\mathbf{B}=\begin{bmatrix}
        1+\gamma & \gamma \phi^*+\xi^*\\ \gamma\phi+\xi & 1+\gamma
    \end{bmatrix}.\] Nevertheless, the corresponding $\bmatext$ constructed with (\ref{eq: B_extended_definition}) does not satisfy the conditions of Corollary~\ref{coro: linearity} in general. Therefore, even for the simplest possible configuration ($n=2$)  with {\it complex channel and noise covariance matrices} and arbitrary $\gamma$, the corresponding $\esthmmse$ is generally a nonlinear function of the quantized observation vector $\r$, as shown in the following proposition. 
\begin{proposition}
    \label{prop: complex_cov_2_by_2}
    Let $n=2$ with the  complex channel and noise correlation matrices  given by \eqref{eq: Phi_Xi_equal_corr}. Then the  corresponding MMSE channel estimator admits
    \begin{align}
        \esthmmse(\r) = s^*\sqrt{\frac{\gamma}{4\pi(1+\gamma)}}\frac{\boldsymbol{\Phi}}{\Prob{\r}}\left(\frac{\r}{8} + \frac{\mathbf{u}(\r)}{4\pi}\right), \label{eq: 2x2_h_MMSE}
    \end{align}
where the entries of the vector $\mathbf{u}(\r)\in\mathbb{C}^2$ are given by
\begin{align}
u_1(\r)&\!=\!r_2^*\real{r_1}\!
\imag{r_1}\!\left(j\kappa_{1}\!-\!\kappa_{2}\!\right)\!-\!r_1^*\real{r_2}\!
\imag{r_2}\!\kappa_{3}, \label{eq: 2x2_num1}
\\u_2(\r)&\!=\!r_1^*\real{r_2}\!
\imag{r_2}\!\left(\kappa_{2}\!+\!j\kappa_{1}\!\right)\!+\!r_2^*\real{r_1}\!
\imag{r_1}\!\kappa_{3},\label{eq: 2x2_num2}
\end{align}
and 
\begin{align}
    \Prob{\r}&=\frac{1}{16} + \frac{\kappa_{4}}{8\pi}\real{r_1^* r_2} +\frac{\kappa_{5}}{8\pi}\imag{r_1^* r_2} \nonumber
    \\&\phantom{=}+\frac{\left(\mathcal{I}_1-\mathcal{I}_2\right)}{4\pi^2}\real{r_1}\real{r_2}\imag{r_1}\imag{r_2}. \label{eq: 2x2_denom}
\end{align}
Moreover, the specific constants  are as follows: \begin{align*}
\kappa_1&=\arcsin\left(\frac{\real{\beta_{\rm c}}}{\sqrt{1-(\imag{\beta_{\rm c}})^2}}\right), \\\kappa_{2} &= \arcsin\left(\frac{\imag{\beta_{\rm c}}}{\sqrt{1-(\real{\beta_{\rm c}})^2}}\right),
\\\kappa_{3}& = \arcsin(\sin \kappa_1 \sin\kappa_2),
\\\kappa_{4}& =\arcsin(\real{\beta_{\rm c}}), \quad \kappa_{5}=\arcsin(\imag{\beta_{\rm c}}),
\\\mathcal{I}_1 &=\int_0^{|\kappa_{4}|}\arcsin\left(\sqrt{\frac{|\beta_{\rm c}|^2\cos^{2}\theta-(\imag{\beta_{\rm c}})^2}{\cos^2\theta-(\imag{\beta_{\rm c}})^2}}\right)\intd\theta
\end{align*}
and
\begin{align*}
\mathcal{I}_2 
&=\int_0^{|\kappa_{5}|}\arcsin\left(\sqrt{\frac{|\beta_{\rm c}|^2\cos^2\theta-(\real{\beta_{\rm c}})^2}{\cos^2\theta-(\real{\beta_{\rm c}})^2}}\right)\intd\theta
\end{align*}
with 
\begin{align}
    \beta_{\rm c}=\displaystyle \frac{\gamma \phi+\xi}{1+\gamma}\in\mathbb{C}. \label{eq: beta_c}
\end{align}
Consequently, the associated MMSE takes the form 
\begin{align}
\hspace{-1mm} \mathrm{MMSE}_2(\gamma, \phi, \xi)\!
=\!1\!-\!\frac{\gamma}{8\pi(\!1\!+\!\gamma\!)}\!\!\sum_{\r\in\{\pm1\pm\! j\}^2}\!\!\!\!\!\!\frac{\left\lVert\boldsymbol{\Phi}\!\!\left(\!\frac{\r}{8}\!+\! \frac{\mathbf{u}(\r)}{4\pi}\!\right)\!\right\rVert^2}{\Prob{\r}}.
\label{eq: MMSE_2x2_final_form}
\end{align}
\end{proposition}
\begin{IEEEproof}
    The proof leverages Theorem~\ref{theo:general} and the evaluation of a $4\times 4$ normal orthant probability as given in \cite{Childs-1967}. 
\end{IEEEproof}
A careful examination of Proposition~\ref{prop: complex_cov_2_by_2} shows that the MMSE estimator in \eqref{eq: 2x2_h_MMSE} depends on the noise correlation parameter $\xi$ exclusively through the scalar quantity $\beta_{\rm c}$ (c.f. \eqref{eq: beta_c}), which
plays a pivotal role in characterizing the influence of noise correlation. More significantly, it 
yields 
specific non-trivial cases in which the estimator remains {\it linear} even though both correlation parameters are non-zero (i.e., $\phi\neq 0$ and $\xi\neq 0$). To the best of our knowledge, this  noise correlation induced linearity has not been previously reported in the literature. 

To be precise, consider the specific conditions that either $\real{\beta_{\rm c}}=0$
 (i.e., $\real{\gamma \phi+\xi}=0$) or $\imag{\beta_{\rm c}}=0$ (i.e., $\imag{\gamma\phi+\xi}=0$). In particular, corresponding to the condition $\real{\gamma \phi+\xi}=0$, one can determine $\gamma_1$ such that $\gamma_1= -\frac{\real{\xi}}{\real{\phi}}>0$, provided that  $\real{\phi}\neq 0$ with $\real{\xi}$ and $\real{\phi}$ having opposite signs. After some algebraic manipulations,  the MMSE estimator here is shown to assume  the linear form
\begin{align*}
&\esthmmse(\r)\!=s^*\sqrt{\frac{\gamma_1}{\pi(1+\gamma_1)}}\boldsymbol{\Phi}\!\begin{bmatrix}
            1  & \frac{-j2\kappa_2}{\pi}
            \\ \frac{j2\kappa_2}{\pi} & 1           \end{bmatrix}^{-1}\!\r,
    \end{align*}
    where $\kappa_2=\arcsin\left(\frac{\imag{\gamma_1\phi+\xi}}{1+\gamma_1}\right)$. The associated MMSE is given by \[\mathrm{MMSE}_2(\gamma_1, \phi, \xi)= 1-\frac{2\pi\gamma_1\left(1+|\phi|^2-\frac{4\kappa_2}{\pi}\imag{\phi}\right)}{(1+\gamma_1)\left(\pi^2-4\kappa_2^2\right)}.\] On the other hand, corresponding to the condition $\imag{\gamma \phi+\xi}=0$, one can determine $\gamma_2$ such that $\gamma_2= -\frac{\imag{\xi}}{\imag{\phi}}>0$, subject to $\imag{\phi}\neq 0$ with $\imag{\xi}$ and $\imag{\phi}$ having opposite signs. In this case, after some algebra,  the MMSE estimator again assumes the linear form
\begin{align}
        \esthmmse(\r)=s^*\sqrt{\frac{\gamma_2}{\pi(1+\gamma_2)}}\boldsymbol{\Phi}
        \begin{bmatrix}
            1  & \frac{2\kappa_1}{\pi}
            \\\frac{2\kappa_1}{\pi} & 1   
        \end{bmatrix}^{-1}\r,  \label{eqn: linearity_imag}
    \end{align}
where $\kappa_1=\arcsin\left(\frac{\real{\gamma_2\phi+\xi}}{1+\gamma_2}\right)$. The corresponding MMSE is given by \[\mathrm{MMSE}_2(\gamma_2, \phi, \xi)= 1-\frac{2\pi\gamma_2\left(1+|\phi|^2-\frac{4\kappa_1}{\pi}\real{\phi}\right)}{(1+\gamma_2)\left(\pi^2-4\kappa_1^2\right)}.\]
Clearly, the extension of linearity beyond the trivial uncorrelated case $\phi=\xi=0$ is made possible by the additional degree of freedom provided by the noise correlation. Furthermore, it can be verified that, in all cases of linear $\esthmmse$, the condition specified in Corollary~\ref{coro: linearity} is satisfied.

The asymptotic SNR behavior of the estimator in (\ref{eq: 2x2_h_MMSE}) is also governed by the limit of the parameter $\beta_{\rm c}$. To be specific, noting that, as $\gamma\to\infty$, we get $\beta_{\rm c}=\phi+o(1)$, it can be concluded that in the high-SNR regime the noise correlation $\xi$ does not affect the MMSE estimator. Consequently, the corresponding asymptotic MMSE depends only on $\phi$. For $\phi=0$, we have $\beta_{\rm c}\to 0$ as $\gamma\to\infty$, which leads to
\[{\displaystyle \lim_{\gamma\to\infty} \text{MMSE}_2}(\gamma,0,\xi)=1-\frac{2}{\pi},\] 
independently of $\xi$. This confirms that the asymptotic limit $1-\frac{2}{\pi}$, previously known only for $\xi = 0$, also holds under the more general condition of correlated noise ($\xi \neq 0$).
\subsection{Arbitrary $n$ with Real Correlation Structures}
\label{sec: equal-corr}
We now analyze the SIMO system described by \eqref{eq: SIMO_UM_system_model}–\eqref{eq: B_mat_SIMO} for arbitrary $n$ under the real-valued channel and noise correlation structures defined in \eqref{eq: real_Phi_Xi_equal_corr}.
\begin{proposition}
\label{prop: general_N_real_corr_without_whitening}
The MMSE channel estimator for the SIMO channel with real correlation structures as in  \eqref{eq: real_Phi_Xi_equal_corr} is given by
\begin{align}
\label{eq: main_h-MMSE_UW_prop1}
\esthmmse(\r)&=s^*\sqrt{\frac{\gamma}{4\pi(1+\gamma)}}\boldsymbol{\Phi}\mathbf{v}(\r),
    \end{align}
    where, for  $k=1, \ldots, n$, the real and imaginary parts of the $k$-th entry of the vector $\mathbf{v}(\r)\in\mathbb{C}^{n}$  are respectively given by 
    \[\real{v_k(\r)}
       \!=\!\real{r_k}\frac{\mathbb{E}_Z\left\{\!\prod\limits_{\substack{i=1, i\neq k}}^{n}\!\!Q \left(\real{r_i}\!\sqrt{\beta}\;Z\right)\!\!\right\}}{\mathbb{E}_Z\left\{\!\prod\limits_{\substack{i=1}}^{n}\!\!Q\left(\real{r_i}\!\sqrt{
       \frac{\beta}{1-\beta}
       } \;Z\right)\!\!\right\}}\] and  \[\imag{\!v_k(\r)}\!=\!\imag{r_k}\frac{\mathbb{E}_Z\left\{\!\prod\limits_{\substack{i=1, i\neq k}}^{n}\!\!Q\left(\imag{r_i}\!\sqrt{\beta}\;Z\right)\!\!\right\}}{\mathbb{E}_Z\left\{\!\prod\limits_{\substack{i=1}}^{n}\!\!Q\left(\!\imag{r_i}\!\sqrt{\frac{\beta}{1-\beta}} \;Z\right)\!\!\right\}},\] with
        $Z\!\sim\!\mathcal{N}(0, 1)$ and $\beta =\frac{\gamma\phi+\xi}{1+\gamma}\geq 0$.
\end{proposition}
 Proposition~\ref{prop: general_N_real_corr_without_whitening} yields an expression for $\esthmmse(\r)$
  that can be evaluated efficiently via numerical integration, requiring only a single integration variable 
per term (see, e.g., \cite{Tong_MVN_book}). Similar to $\beta_{\rm c}$ in \eqref{eq: beta_c}, here $\beta$ plays a critical role in characterizing the influence of $\xi$. 
A notable special case occurs for $\beta=0$ and $\phi \neq 0$, which corresponds to $\phi$ and $\xi$ having opposite signs. By selecting $\widetilde{\gamma} = -\xi/\phi > 0$, the estimator reduces to the simple linear form
   \begin{align*}
\esthmmse(\r)&=s^*\sqrt{\frac{\widetilde{\gamma}}{\pi(1+\widetilde{\gamma})}}\boldsymbol{\Phi}\r, 
    \end{align*} 
    which is consistent with Corollary~\ref{coro: linearity}. 
    The corresponding MMSE  is given  by  
       \begin{align} \mathrm{MMSE}_n\!\left(\widetilde{\gamma}, \phi, \xi\right)\!=\!1\!-\!\frac{2\widetilde{\gamma}(1\!+\!(n\!-\!1)\phi^2)}{\pi(1+\widetilde{\gamma})}, \label{eq: MMSE_realB_interplay}
    \end{align}
    with  $\phi, \xi\in\big(\!-\frac{1}{n-1}, 0\big)\cup (0,1)$.  The observed expansion of the linear region is induced by noise correlation. However, as $n \to \infty$, this extended linear region vanishes, since the interval $\big(-\frac{1}{n-1}, 0\big)\cup(0,1)$ converges to $(0, 1)$.
    Moreover, the condition $\beta = 0$ is trivially
    satisfied for $\phi = \xi = 0$. Consequently, the estimator reduces to $\esthmmse(\mathbf{r}) = s^* \sqrt{\frac{\gamma}{\pi(1+\gamma)}} \, \mathbf{r}$, with the corresponding MMSE given by $\mathrm{MMSE}_n(\gamma, 0, 0) = 1 - \frac{2\gamma}{\pi(1+\gamma)}$, in agreement with \cite{Fesl_etal_TSP2023, Ding_et_al_2025}. In contrast, for $n=2$, the condition of Corollary~\ref{coro: linearity} is  
    satisfied for all $\gamma>0$, $\xi, \phi\in(-1, 1)$, and thus  the estimator in 
    \eqref{eq: main_h-MMSE_UW_prop1} simplifies to the linear form
    \begin{align}
    \esthmmse(\r)=s^*\!\sqrt{\frac{\gamma}{\pi(1+\gamma)}}\boldsymbol{\Phi}\!\begin{bmatrix}
        1&\frac{2\arcsin\beta}{\pi}\\\frac
        {2\arcsin\beta}{\pi}&1
    \end{bmatrix}^{-1}\!\r, \label{eq: real_2x2_equal_corr}
\end{align}
    which reduces to the AWGN result in \cite[(63)]{Ding_et_al_2025} for $\xi=0$.

    Asymptotically, following the limit $\displaystyle \lim_{\gamma\to \infty }\beta=\phi$, we conclude that the noise correlation $\xi$ has a  negligible impact on the MMSE estimator. Consequently, the asymptotic MMSE depends only on $\phi$. In the special case of $\phi = 0$, we obtain
    $ {\displaystyle\lim_{\gamma\to\infty}\mathrm{MMSE}_n(\gamma, 0, \xi)}= 1-\frac{2}{\pi}$,  regardless of the values of $\xi$ and $n$.  This further generalizes the asymptotic limit \(1-\frac{2}{\pi}\) previously known only for the AWGN case ($\xi = 0$) to the case of correlated noise ($\xi \neq 0$). Similar observations have been made for $n=2$ with complex-valued $\phi$ and $\xi$ in Section~\ref{sec: 2x2_cov_complex}.
    
 \section{Numerical Results}
 \begin{figure}[t]
\centering
\begin{subfigure}[T]{0.241\textwidth}
  \resizebox{\linewidth}{!}{  \input{./fig_1a.tex}}
    \caption{MMSE vs.  $\gamma$ for $\arg(\phi)\!=\!-\frac{\pi}{6}$ and $\arg(\xi)\!=\!-\frac{\pi}{3}$.}
    \label{fig1: fig_1a}
\end{subfigure}
\hfill
\begin{subfigure}[T]{0.241\textwidth}
\resizebox{\linewidth}{!}{
    \input{./new_fig_1b.tex}}
    \caption{MMSE vs. $(\arg(\phi), \arg(\xi))$ for  $|\phi|\!\!=\!0.9$, $|\xi|\!\!=\!0.8$, and $\gamma\!=\!5$ dB.}
    \label{fig1: fig_1b}
\end{subfigure}     
\caption{Illustration of the parametric dependence of MMSE in a SIMO system with $n=2$, highlighting the effects of SNR $\gamma$ as well as complex-valued channel correlation $\phi$ and noise correlation $\xi$.}
\label{fig: fig1}
\end{figure}
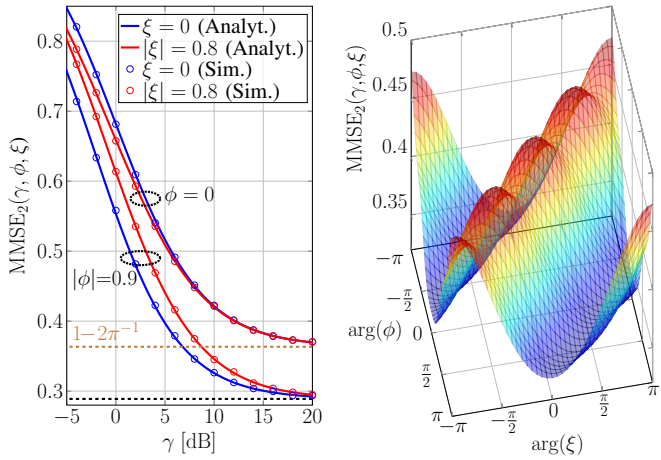
\begin{figure}[t]
\centering
\input{./fig_2.tex}
    \caption{MMSE versus $\gamma$ for a SIMO system with $n=16$, evaluated for different values of $\xi\in\{0, 0.5, 0.9\}$ and $\phi\in\{0,0.9\}$. }
\label{fig: fig2}
\end{figure}
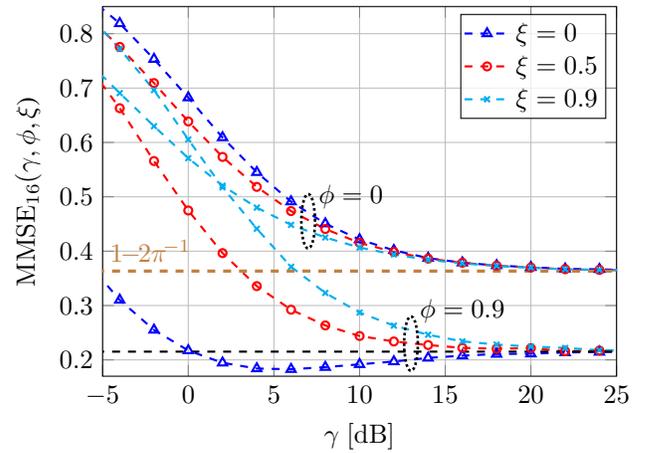    
   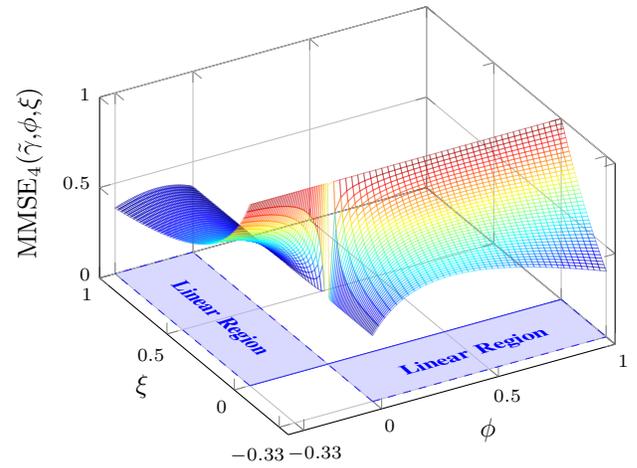
\begin{figure}[t]
\centering
\input{./fig_3D.tex}
\caption{Illustration of the extended linearity region induced by noise correlation $\xi$ and the corresponding MMSE for a SIMO system with $n=4$ (see \eqref{eq: MMSE_realB_interplay}).}
\label{fig: fig_3D}
\end{figure}

This section presents numerical simulations to validate and extend the above analytical findings.
We begin with a SIMO system using two receive antennas (i.e., $n=2$) under the complex correlation structures defined in \eqref{eq: Phi_Xi_equal_corr}. Figure~\ref{fig1: fig_1a} plots the MMSE versus the transmit SNR $\gamma$ for parameter values $\xi=|\xi|e^{-\frac{j\pi}{3}}$,  $|\xi|\in\{0, 0.8\}$ and $\phi=|\phi|e^{-\frac{j\pi}{6}}$, $|\phi|\in\{0, 0.9\}$. It is observed that the analytical results derived from \eqref{eq: 2x2_denom} and \eqref{eq: MMSE_2x2_final_form} show excellent agreement with Monte Carlo simulations based on \eqref{eq: MMSE_SIMO_general}--\eqref{eq: 2x2_h_MMSE}.
As can be seen from the graph, for uncorrelated channels (i.e., $\phi=0$), increasing the noise correlation reduces the MMSE at low-to-medium SNR, confirming its beneficial effect in this regime. Conversely, under strong channel correlation (e.g., $|\phi|=0.9$), the MMSE increases with noise correlation. Moreover, as $\gamma$ increases, the asymptotic MMSE becomes increasingly sensitive only to $\phi$, confirming that the noise correlation $\xi$ does not affect the asymptotic performance.  These results highlight a non-uniform, context-dependent impact, where the influence of noise correlation is dictated by the underlying channel correlation.

Given the complex-valued nature of the correlations, we now examine the dependence of MMSE on the phase components 
$\arg(\xi)\in(-\pi,\pi]$ and $\arg(\phi)\in(-\pi,\pi]$ corresponding to the parameters $|\phi|=0.9$, $|\xi|=0.8$, and $\gamma=5 $ dB. This dependence, shown in Fig.~\ref{fig1: fig_1b}, reveals that specific phase combinations yield lower MMSE than others, indicating optimal operating regions.

To examine scalability, we analyze a relatively larger SIMO configuration with $n=16$ receive antennas under the real correlation framework of \eqref{eq: real_Phi_Xi_equal_corr}. Figure~\ref{fig: fig2} illustrates the MMSE as a function of $\gamma$ for parameter values $\xi \in \{0, 0.5, 0.9\}$ and $\phi \in \{0, 0.9\}$. Reaffirming the trend from Fig.~\ref{fig1: fig_1a},  noise correlation is beneficial (i.e., reducing the MMSE) in the absence of channel correlation ($\phi=0$), but becomes detrimental when channels are highly correlated ($\phi=0.9$).

Finally, Fig.~\ref{fig: fig_3D} displays  the enhanced linear estimation region enabled by noise correlation $\xi$ as well as the corresponding MMSE performance, evaluated via \eqref{eq: MMSE_realB_interplay} with $\widetilde{\gamma}=-\xi/\phi$ and $\phi\neq 0$ for a SIMO system with $n=4$ antennas. Remarkably, Fig.~\ref{fig: fig_3D} reveals that   this particular linear MMSE estimator 
performs
most effectively when 
$\xi$ is high and $\phi$ is negative, or when $\phi$ is high and $\xi$ is negative.
This clearly demonstrates the utility of the additional degree of freedom offered by noise correlation $\xi$.  In contrast, for the AWGN case with $\xi=0$, the linear region disappears, which aligns with the established nonlinear nature of $\esthmmse$ with correlated channels
($\phi\neq 0$). 
However, when  $\xi=\phi=0$, $\esthmmse$ is  again linear   for all $\gamma>0$. 

\section{Conclusions}
This paper develops a unified framework for MMSE channel estimation in one-bit MIMO systems with noise correlation. It provides a general estimator that generalizes prior work, establishes linearity conditions, and gives efficient SIMO forms. The analyses reveal that noise correlation can extend the estimator’s linear region when channels are also correlated,
while simulations show it improves performance for uncorrelated channels at low-to-medium SNR but degrades it for strongly correlated channels. These new insights offer important design guidance for communication systems employing one-bit ADCs.

\bibliographystyle{IEEEbib}
\balance
\bibliography{IEEEabbr, ./isit26}
\end{document}

%% file: symbols.tex
\def\a{\ensuremath{\mathbf{a}}}
\def\n{\ensuremath{\mathbf{n}}}

\def\r{\ensuremath{\mathbf{r}}}
\def\Y{\ensuremath{\mathbf{Y}}}
\def\y{\ensuremath{\mathbf{y}}}

\def\h{\ensuremath{\mathbf{h}}}

\def\I{\ensuremath{\mathbf{I}}}
\def\1{\ensuremath{\mathbf{1}}}
\def\0{\ensuremath{\mathbf{0}}}

\def\bmat{\mathbf{B}}
\def\bmatext{\mathbf{B}_{\rm c}}
\def\esthmmse{\widehat{\h}_{\mathrm{mmse}}}

\def\intd{\mathrm{d}}
\def\trace{\mathrm{tr}}

\newcommand\real[1]{\Re\left\{#1\right\}}
\newcommand\imag[1]{\Im\left\{#1\right\}}
\newcommand\Prob[1]{\mathrm{Pr}\!\left(#1\right)}
\newcommand\OrthP[1]{\mathcal{P}\!\left[#1\right]}
\newcommand\myexp[1]{\ensuremath{\mathbb{E}\left\{#1\right\}}}
\newcommand{\sgn}{\mathrm{sgn}}
\def\vec{\ensuremath{\mathrm{vec}}}
\newcommand\quantoneb[1]{\ensuremath{\mathcal{Q}_{\mathrm{1b}}\left(#1\right)}}
\newcommand\diag[1]{\ensuremath{\mathrm{diag}\left(#1\right)}}

\newcommand{\Herm}{^H}
\newcommand{\Trans}{^T}


%% file: fig_1a.tex
\begin{tikzpicture}

\begin{axis}[
	width=8cm,
	height=13cm,
    scale only axis,
	xmin=-5, xmax=20,
	ymin=0.28, ymax=0.85,
	xlabel={$\gamma \;\mathrm{[dB]}$},
	ylabel={$\mathrm{MMSE}_2(\gamma,\phi,\xi$)},
	ytick={0.3, 0.4, 0.5,0.6, 0.7, 0.8},
	xlabel near ticks,
	ylabel near ticks,
	x label style={font=\fontsize{20}{28}\selectfont},
	y label style={font=\fontsize{20}{28}\selectfont},
	ticklabel style={font=\fontsize{20}{28}\selectfont},
	legend style={at={(0.98,0.98)}, anchor=north east},
	legend style={font=\fontsize{20}{28}\selectfont, inner sep=1pt, fill opacity=0.75, draw opacity=1, text opacity=1},
	legend cell align=left,
	grid=both,
	title style={font=\scriptsize},
]

\addplot[line width=2pt, blue]
table [x=Var1, y=Var2, col sep=comma] {./n2MMSE_no_corr.txt};
\addlegendentry{$\xi=0$ (Analyt.)};

\addplot[line width=2pt, red]
table [x=Var1, y=Var2, col sep=comma] {./n2MMSE_noise_corr_only.txt};
\addlegendentry{$|\xi|=0.8$ (Analyt.)};

\addplot[line width=2pt, blue, forget plot]
table [x=Var1, y=Var2, col sep=comma] {./n2MMSE_ch_corr_only.txt};

\addplot[line width=2pt, red, forget plot]
table [x=Var1, y=Var2, col sep=comma] {./n2MMSE_ch_noise_corr_both.txt};

\addplot[thick, blue, only marks, mark=o,  mark size=3pt]
table [x=Var1, y=Var2, col sep=comma] {./sim2_MMSE_No_cor.txt};
\addlegendentry{$\xi=0$ (Sim.)};

\addplot[thick, red, only marks, mark=o,  mark size=3pt]
table [x=Var1, y=Var2, col sep=comma] {./sim2_MMSE_Ncorr_only.txt};
\addlegendentry{$|\xi|=0.8$ (Sim.)};

\addplot[thick, blue, only marks, mark=o,  mark size=3pt]
table [x=Var1, y=Var2, col sep=comma] {./sim2_MMSE_ch_cor_only.txt};

\addplot[thick, red, only marks, mark=o,  mark size=3pt]
table [x=Var1, y=Var2, col sep=comma] {./sim2_MMSE_ch_N_cor_both.txt};

\addplot[line width = 2pt, brown, dashed]
table [x=Var1, y=Var2, col sep=comma] {./const_1m2pi.txt};

\addplot[line width = 2pt, black, dashed]
table [x=Var1, y=Var2, col sep=comma] {./const_othern2.txt};

\draw[densely dotted, line width=2pt] (3,0.575) ellipse (1.5 and 0.01);

\draw[densely dotted, line width=2pt] (2.5,0.49) ellipse (2 and 0.01);


\node[black,  font=\fontsize{20}{28}\selectfont] at (7.5,0.58) {$\phi = 0$};

\node[black, font=\fontsize{20}{28}\selectfont] at (-1.15,0.46) {$|\phi| {=} 0.9$};

\node[brown, font=\fontsize{20}{28}\selectfont] at (-1,0.385) {$1\!\!-\!2\pi^{-1}$};

\end{axis}

\end{tikzpicture}

%% file: new_fig_1b.tex
\begin{tikzpicture}
\newcommand{\viewa}{-30}
\newcommand{\viewb}{30}

\newcommand{\viewc}{58}
\newcommand{\viewd}{30}

\begin{axis} [
    xtick = {-pi, -pi/2, 0, pi/2, +pi}, 
    xticklabels = {$-\pi$, $-\frac{\pi}{2}$, $0$, $\frac{\pi}{2}$, $\pi$},
    ytick = {-pi, -pi/2, 0, pi/2, pi},
    yticklabels = {$-\pi$, $-\frac{\pi}{2}$, $0$, $\frac{\pi}{2}$, $\pi$},
    ztick = {0.35, 0.4, 0.45, 0.5},
    width=8cm,
	height=13.6cm,
    scale only axis,
    xmax=pi,
    xmin=-pi,
    ymax=pi,
    ymin=-pi,
    zmax=0.50,
    zmin=0.32,
    view={78.5}{39},
    colormap/jet,
     every axis x label/.style={at={(ticklabel cs:0.2)}, anchor=north, xshift=-15pt},  
     xticklabel style ={xshift=-5pt, yshift=10pt},
    xlabel = $\arg(\phi)$, ylabel = $\arg(\xi)$, zlabel= $\mathrm{MMSE}_2(\gamma{,}\phi{,}\xi)$,
    x label style={font=\fontsize{20}{28}\selectfont},
	y label style={font=\fontsize{20}{28}\selectfont},
    z label style={font=\fontsize{20}{28}\selectfont},
    ticklabel style = {font = \fontsize{20}{28}\selectfont},
	grid,
    mesh/rows=41,  
    mesh/cols=41, 
]

\addplot3[surf, mesh/ordering=colwise, opacity=0.4] table {./MMSE_n2_phase3D.dat};
       
\end{axis}
\end{tikzpicture}

%% file: fig_2.tex
\begin{tikzpicture}

\begin{axis}[
	width=0.95\linewidth,
	height=6.5cm,
	xmin=-5, xmax=25,
	ymin=0.17, ymax=0.85,
	xlabel={$\gamma \;\mathrm{[dB]}$},
	ylabel={$\mathrm{MMSE}_{16}(\gamma,\phi,\xi$)},
	ytick={0.2, 0.3, 0.4, 0.5,0.6, 0.7, 0.8, 0.9},
	xlabel near ticks,
	ylabel near ticks,
	x label style={font=\fontsize{10}{14}\selectfont},
	y label style={font=\fontsize{10}{14}\selectfont},
	ticklabel style={font=\fontsize{10}{14}\selectfont},
	legend style={at={(0.98,0.98)}, anchor=north east},
	legend style={font=\fontsize{10}{14}\selectfont, inner sep=1pt, fill opacity=0.75, draw opacity=1, text opacity=1},
	legend cell align=left,
	grid=both,
	title style={font=\scriptsize},
]

\addplot[line width=0.8pt, blue, dashed,  mark = triangle, mark options=solid, mark size=2pt]
table [x=Var1, y=Var2, col sep=comma] {./mmse100_vec.txt};
\addlegendentry{$\xi=0$};

\addplot[line width=0.8pt,
red, dashed, mark=o, mark options=solid, mark size=1.5pt]
table [x=Var1, y=Var2, col sep=comma] {./mmse105_vec.txt};
\addlegendentry{$\xi=0.5$};

\addplot[line width=0.8pt,
cyan, dashed, mark=x,
            mark options=solid, mark size=1.5pt]
table [x=Var1, y=Var2, col sep=comma] {./mmse109_vec.txt};
\addlegendentry{$\xi=0.9$};

\addplot[line width=0.8pt,
blue, dashed, mark=triangle, mark options=solid, mark size = 2pt]
table [x=Var1, y=Var2, col sep=comma] {./ch09_mmse100_vec.txt};

\addplot[line width=0.8pt,
red,  dashed, mark=o,
            mark options=solid, mark size = 1.5pt]
table [x=Var1, y=Var2, col sep=comma] {./ch09_mmse105_vec.txt};

\addplot[line width=0.8pt,
cyan,  dashed, mark=x,
            mark options=solid, mark size = 1.5pt]
table [x=Var1, y=Var2, col sep=comma] {./ch09_mmse109_vec.txt};

\addplot[line width = 1.2pt, brown, dashed]
table [x=Var1, y=Var2, col sep=comma] {./const_1m2pi.txt};

\addplot[line width = 0.8pt, black, dashed]
table [x=Var1, y=Var2, col sep=comma] {./const_othern16_ch09.txt};

\draw[densely dotted, line width = 1pt] (7,0.455) ellipse (0.4 and 0.05);

\draw[densely dotted, line width=1pt] (13,0.23) ellipse (0.4 and 0.05);

\node[black, font=\fontsize{10}{14}\selectfont] at (9.5,0.5) {$\phi = 0$};

\node[black, font=\fontsize{10}{14}\selectfont] at (16,0.29) {$\phi = 0.9$};

\node[brown, font=\fontsize{10}{14}\selectfont] at (-2.2,0.398) {$1\!\!-\!\!2\pi\!^{-1}$};

\end{axis}

\end{tikzpicture}

%% file: fig_3D.tex
\begin{tikzpicture}
\newcommand{\viewa}{-30}
\newcommand{\viewb}{30}

\newcommand{\viewc}{58}
\newcommand{\viewd}{30}
\pgfplotsset{
    /pgfplots/colormap={new2}{[1cm]rgb255(0cm)=(0,50,200) rgb255(3cm)=(200,155,0)
        rgb255(6cm)=(255,0,255) rgb255(8cm)=(255,0, 0)}
}
\begin{axis} [
    xtick = {-1/3, 0, 0.5,1},
    ytick = {-1/3,0,0.5,1},
xmax=1,
xmin=-0.4,
ymax=1,
ymin=-0.4,
    zmax=1,
    zmin=0,
    view={-30}{45},
     every axis x label/.style={at={(ticklabel cs:0.7)}, anchor=north, xshift=-20pt, yshift=5pt}, 
      every axis y label/.style={at={(ticklabel cs:0.8)}, anchor=north, xshift=+18pt},
    xlabel = $\phi$, ylabel = $\xi$, zlabel= $\mathrm{MMSE}_4(\tilde{\gamma}{,}\phi{,}\xi)$,
    ticklabel style = {font = \scriptsize},
	grid,
    colormap/jet
]

\addplot3 [mesh, opacity=0.4, domain=0.0001:1, y domain=-1/3:0, samples=40] 
    	{1-2*y*(1+3*x*x)/((y-x)*pi)};
        
      \addplot3 [mesh, opacity=0.4, domain=-1/3:-0.001, y domain=0:1, samples=45] 
    	{1-2*y*(1+3*x*x)/((y-x)*pi)};

\draw [ dashed, blue!] (-1/3,0) -- (-1/3,1);
\draw [ dashed, blue!] (0,-1/3) -- (1,-1/3);
\draw [ blue!] (-1/3,0) -- (1,0);
\draw [ dashed, blue!] (0,-1/3) -- (0,1);
\draw [dashed, blue!] (1,0) -- (1,-1/3);
\draw [dashed, blue!] (0,1) -- (-1/3,1);

    \addplot3[draw=none
          ]
    coordinates {
        (0,0,0)
        (1,0,0)
        (1,-1/3,0)
        (0, -1/3, 0)
    };
\fill[blue!30,opacity=0.5] (0,0,0) -- (1,0,0) -- (1,-1/3,0) -- (0, -1/3, 0) -- cycle;

\addplot3[draw=none
          ]
    coordinates {
        (0,0,0)
        (-1/3,0,0)
        (-1/3,1,0)
        (0, 1, 0)
    };
\fill[blue!30,opacity=0.5] (0,0,0) -- (-1/3,0,0) -- (-1/3,1,0) -- (0, 1, 0) -- cycle;

\node[color=blue,
        cm={cos(\viewa),-sin(\viewa)*sin(\viewb),
        sin(\viewa),cos(\viewa)*sin(\viewb),
        (0,0)}
            ]   
     at (axis cs:0.75,0.5,-0.9) {\bf Linear Region}; 

\node[color=blue,
        cm={cos(\viewc),-sin(\viewc)*sin(\viewd),
        sin(\viewc),cos(\viewc)*sin(\viewd),
        (0,0)}
            ]   
     at (axis cs:0.06,1.2,-0.9) {\bf Linear Region};

\end{axis}
\end{tikzpicture}